\documentclass[prl,twocolumn,groupedaddress,10pt]{revtex4-1}
\usepackage[pdftex,bookmarks=true,bookmarksnumbered=true,plainpages=false,pdfpagelabels]{hyperref}

\usepackage{amsmath}
\usepackage{amssymb}
\usepackage{float}
\usepackage{bm,times}
\usepackage{graphicx}
\usepackage{units}
\usepackage{threeparttable} 
\usepackage{comment}

\newcommand{\SiN}{$\rm Si_3N_4$}
\newcommand{\um}{$\mu \rm m$}

\DeclareFontFamily{U}{euc}{}
\DeclareFontShape{U}{euc}{m}{n}{<-6>eurm5<6-8>eurm7<8->eurm10}{}%
\DeclareSymbolFont{AMSc}{U}{euc}{m}{n}
\DeclareMathSymbol{\umu}{\mathord}{AMSc}{"16}

\begin{document}

\title{A phononic bandgap shield for high-$Q$ membrane microresonators}
\author{P.-L. Yu$^{1,2}$}
\author{K. Cicak$^{3}$}
\author{N. S. Kampel$^{1,2}$}
\author{Y. Tsaturyan$^{1,2}$}
\altaffiliation{Current address: QUANTOP, Niels Bohr Institute, University of Copenhagen, Blegdamsvej 17, 2100 Copenhagen, Denmark}
\author{T. P. Purdy$^{1,2}$}
\author{R. W. Simmonds$^{3}$}
\author{C.~A.~Regal$^{1,2}$}
\email{regal@colorado.edu}
\affiliation{$^1$JILA, University of Colorado and National Institute of Standards and Technology}
\affiliation{$^2$Department of Physics, University of Colorado, Boulder, Colorado 80309, USA}
\affiliation{$^3$National Institute of Standards and Technology, Boulder, Colorado 80305, USA }
\date{\today}

\begin{abstract}
A phononic crystal can control the acoustic coupling between a resonator and its support structure. We micromachine a phononic bandgap shield for high $Q$ silicon nitride membranes and study the driven displacement spectra of the membranes and their support structures. We find that inside observed bandgaps the density and amplitude of non-membrane modes are greatly suppressed, and membrane modes are shielded from an external mechanical drive by up to 30~dB.
\end{abstract}

\maketitle
Micro- and nano-mechanical resonators offer great potential for precision sensing and realizing non-classical states of relatively massive objects~\cite{OConnell2010,Chan2011a,Teufel2011,Palomaki2013a}. One promising platform is silicon nitride (\SiN) membrane resonators on silicon substrates~\cite{Verbridge2006,Zwickl2008,Wilson2009}, in which large tensile stress results in a $Q$-frequency product above $10^{13}$~Hz~\cite{Unterreithmeier2010,Schmid2011a,Yu2012c}.  Coupling these high-$Q$ membranes to a Fabry-P\'erot cavity has enabled quantum measurements on macroscopic objects and preparing mechanical modes close to the quantum ground state~\cite{Thompson2008,Purdy2012,Purdy2013a,Purdy2013b}.

Currently, an important limitation to the displacement sensitivity and radiation pressure effects of \SiN~membranes in an optical cavity comes from the coupling between the membrane and the support structure~\cite{Jockel2011,Wilson-Rae2011,Yu2012c,Yu2013}. This coupling results in 1) radiation loss, in which the energy of membrane modes radiate into the substrate, and 2) substrate noise, in which the mechanical modes of the silicon frame limit the optomechanical cooling. Thus far, the radiation loss in these devices has been addressed mainly by varying the techniques for grasping the silicon frame~\cite{Wilson2009,Zwicklthesis,Purdy2012}.  However more sophisticated techniques for control of acoustic waves have been extensively studied in the field of acoustic metamaterials. In particular, a phononic crystal (PnC) with acoustic bandgaps can be used to filter or confine acoustic waves~\cite{Narayanamurti1979,Sigalas1992,Sigalas1992,Kushwaha1993,Martinez-Sala1995,Maldovan2013}. The use of PnC bandgaps to suppress the radiation loss of gigahertz, in-plane resonators has been demonstrated in the field of MEMS and optomechanics~\cite{Mohammadi2009a,Chan2011a}.

\begin{figure}
\begin{center}
\includegraphics[width=0.42\textwidth]{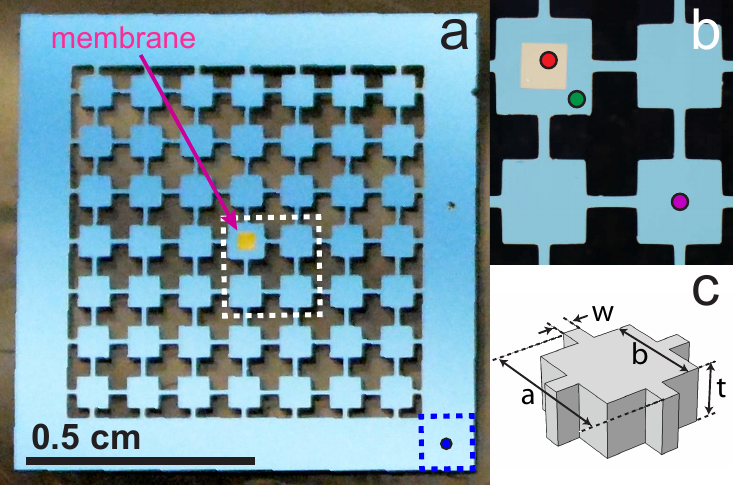}
\caption{(a) Photograph of device A. The outermost chip frame (CF) is connected to the piezoelectric actuator at four corners (the blue dashed region). (b) Expanded view of the white dashed regions shows a square membrane (M, yellow) surrounded by a membrane frame (MF, light blue) and a PnC unit cell (PnC, light blue). The red, green, and purple spots in (b) and the blue spot in (a) are locations of displacement measurements in Fig.~\ref{fig2}~(c)-(f). (c) Schematic of the PnC unit cell and definitions of the geometry parameters. See Table~\ref{table1} for the values of these parameters for the devices A and B.
}\label{fig1}
\end{center}
\end{figure}

\begin{figure*}
\includegraphics[width=1\textwidth]{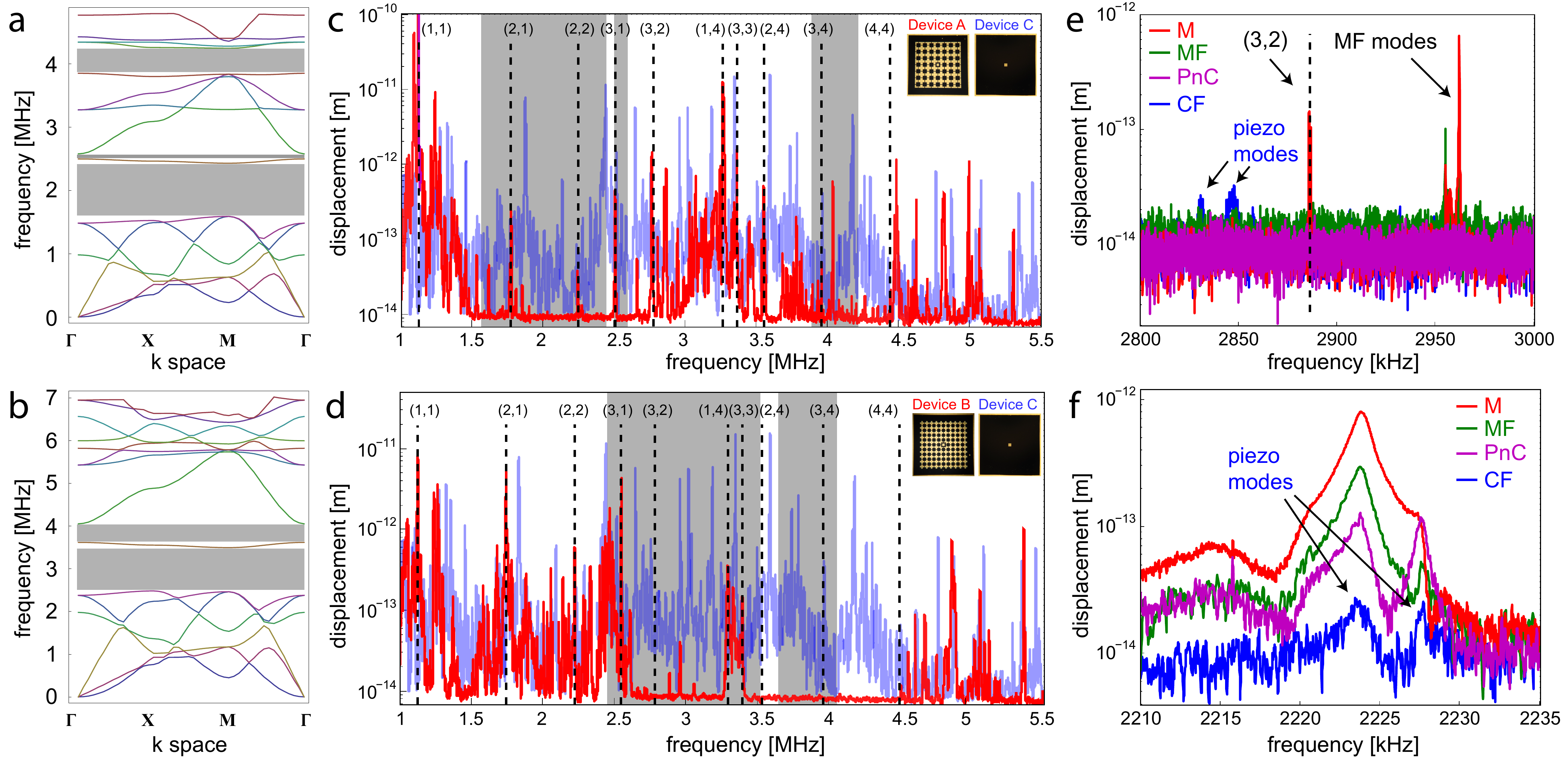}
\caption{ Frequency-dependent mechanical response. (a)-(b) Simulated band diagrams for infinite number of the unit cells used in devices A and B. Bandgap ranges are shown in grey. (c)-(d) Measured membrane displacement spectra of devices A, B, and C. The data are smoothed with a 4~kHz bandwidth. The ranges of ideal bandgaps are shown in grey. Membrane modes predicted based on the observed fundamental mode frequencies of devices A and B (up to the (4,4) mode) are shown by dashed lines. (e)-(f) Probing the non-membrane modes via measuring displacement spectra at different locations on device B:  At the membrane (red), the membrane frame (MF, green), the PnC (purple), and the corner of chip frame (CF, blue). (e) An example spectral region in an observed bandgap in device B. (f) An example of two non-membrane modes of device B outside the observed bandgaps.} \label{fig2}
\end{figure*}

In this work we demonstrate a high-tension membrane inside of a silicon PnC structure that provides a shield for acoustic modes at megahertz frequencies.  We probe the membrane modes and the non-membrane modes by measuring the displacement spectra of the membrane and different components of the support structure. We find that inside the observed bandgaps the density and the amplitude of the non-membrane modes are greatly suppressed. In addition, the membrane modes inside the observed bandgap are shielded from an external mechanical drive by up to 30~dB.

The device consists of a patterned silicon substrate with a center island that contains high-tension square film of \SiN~suspended across a mm-scale frame [Fig.~\ref{fig1}(a)]. The unit cell length scale required to create a bandgap centered at a frequency $f$ can be estimated by $\lambda/2=v/2f\sim\mathrm{1~mm}$, where $\lambda$ and $v$ are the acoustic wavelength and velocity in silicon, respectively. For bandgaps centered at megahertz frequencies, we can fit 3 to 4 unit cells around the membrane with a 1~cm square chip.  Our unit cell is composed of a square block with four bridges [Fig.~\ref{fig1}(c)]~\cite{Safavi2010,Alegre2010}. 

We study two different devices (A and B) with different PnC shields [Fig.~\ref{fig1}(a)], and a reference device (C) without the PnC shield.  (See Table~\ref{table1} for measured geometry parameters.)  In Figs.~\ref{fig2}(a) and (b), we display the band diagrams for the two different PnCs with infinite number of unit cells; this calculation is completed with the finite-element-method (FEM) software COMSOL using the measured device parameters. 

Fabrication of the devices begins with the growth of a 100-nm-thick \SiN~film by LPCVD on both sides of a 300-\um~thick Si wafer.  The membrane and PnC structure are created in two sequential steps; each starts with patterned removal of the back \SiN~layer followed by deep reactive ion etching (DRIE) for bulk Si machining.  In the first step, the DRIE stops 10Õs of microns short of etching fully through the wafer, and a KOH wet etch completes the release of the square \SiN~membrane on the front of the wafer.  In the second step, the PnC crosses are micromachined with DRIE all the way through the wafer (resulting in PnC holes that are vertical  to $\sim1^{\circ}$).  During fabrication (except the KOH step), the front side of the wafer is glued with processing adhesive to a protection substrate, and the final devices are released from the protection substrate and cleaned using solvents and a sulfuric-acid-based solution.

\begin{table}
\begin{threeparttable}
\caption{Measured geometry parameters of the devices}\label{table1}
\begin{tabular}{ l    c  c  c  c  }
    \hline
 Definition [\um]             &   symbol    &Device A            & Device B              & Device C~\cite{Norcada}  \\ \hline \hline
number of unit cells*        &                 &  3                     &  4.5                      & - \\
unit cell size                    &    $a$       & 1100~\um         & 800~\um             & - \\
block length                    &     $b$       & 686                   & 542                     & - \\
bridge width                    &   $w$        & 97                     & 96                       & - \\
wafer thickness               &  $t$           & 300                   & 300                     & - \\
membrane length            &   $l$          & 372                   &  367                     & 500  \\
membrane frame size      &                  & 786                   & 783                     & $10^4$ \\
membrane thickness        & $$             & 0.1                    & 0.1                      & 0.04 \\
    \hline
    \end{tabular}
\begin{tablenotes}
\item [*] between the center and the edge of the chip
\end{tablenotes}
\end{threeparttable}
\end{table}

The membrane resonator vibrates like a drum with discrete frequencies given by $f_{mn}=\sqrt{\sigma(m^2+n^2)/4\rho l^2}$, where $\sigma$ is the tensile stress, $(m,n)$ are integer mode indices representing the number of antinodes, $\rho$ is the volume mass density, and $l$ is the membrane side length. The fabricated membranes in the PnCs are experimentally confirmed to be under a high tensile stress of 1~GPa: The fundamental membrane frequency for devices A and B is 1.1~MHz.

To characterize the mechanical properties of the devices, we excite the chip at different frequencies through a piezoelectric ring actuator connected to the frame corners and measure displacement using a Mach-Zehnder interferometer.  First we present studies in which we probe the displacement of the \SiN~membrane. We position the optical spot slightly off the membrane center to allow a variety of modes to be probed. The driven displacements as a function of frequency for devices A and B are compared with that of a control device C in Fig.~\ref{fig2}(c) and (d), respectively. We find that the displacement is clearly suppressed in the frequency ranges of 1.5-2.75~MHz and 4.05-4.45~MHz (2.65-3.25~MHz and 3.5-4.5~MHz) for device A (B), resulting in a flat response that is limited by the shot noise of optical detection. These ``observed bandgaps'' roughly overlap with the calculated bandgaps [grey regions in both Fig.~\ref{fig2}(a),(b) and (c),(d)]. The center frequencies of the observed and predicted bandgaps are consistent within $\sim$10$\%$.

Most of the modes we see in Fig.~\ref{fig2}(c) and (d) are non-membrane modes; the finite number of membrane modes are shown by dashed lines. Physically the chip consists of (1) the membrane (M), (2) the membrane frame (MF), (3)~the PnC structure (PnC), and (4) the chip frame (CF) [Fig.~\ref{fig1}(a)]. The membrane and the MF together form a ``defect" embedded in the PnC lattice. We optically probe the MF, the PnC, and the CF by focusing on the three different locations indicated in Fig.~\ref{fig1}(b). Looking at these spectra in conjunction with the membrane displacement, we can understand the origin of the non-membrane modes. The piezoelectric actuator itself has frequency-dependent structure, and measuring at the CF reveals the information about this structure. Mainly the displacement measured on the corner of CF is limited by the detection noise, but some ``piezo-modes'' are clearly identifiable [see Fig.~\ref{fig2}(e) and (f) for two examples].

We find that the combined spectra have distinct features inside and outside the observed bandgaps. Inside the observed bandgap, the spectra of the PnC, the MF, and the membrane are flat except a couple of ``defect modes'' observed in the spectra of the MF and the membrane [see Fig.~\ref{fig2}(e) for one example].   While the mechanical modes of the MF cannot be completely avoided in the bandgap, they only occur sparsely and are clearly separable from the membrane modes. Outside the observed bandgaps, most modes except the membrane modes have comparable motion in the membrane, the MF, and the PnC [see Fig.~\ref{fig2}(f) for one example]. We also find that piezo modes greatly enhance the motion of other components, while inside the observed bandgaps the piezo modes do not induce any observed motion of other components [compare Fig.~\ref{fig2}(f) and (e)].

\begin{figure} \begin{center}
\includegraphics[width=0.5\textwidth]{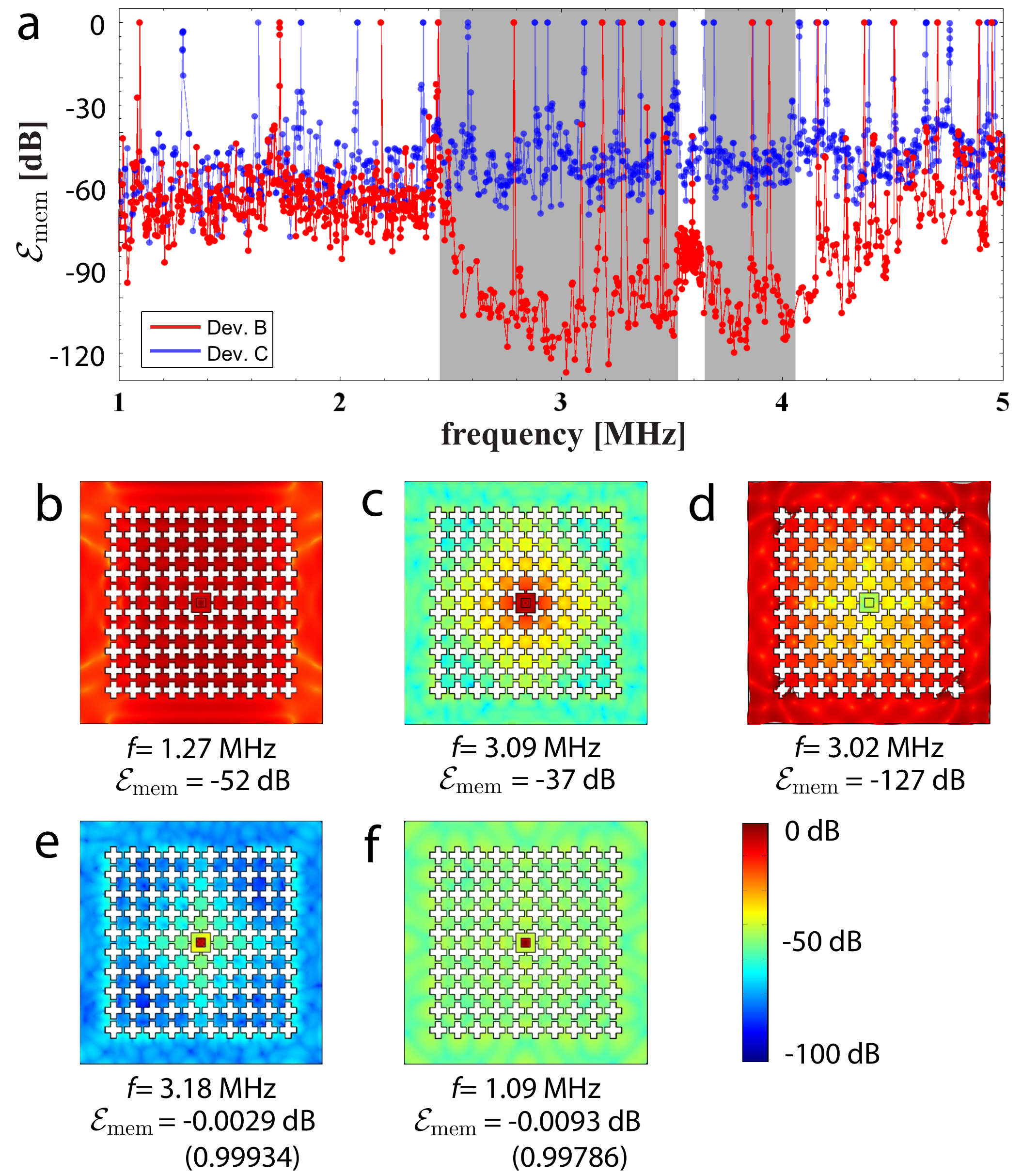}
\caption{Simulated membrane and non-membrane modes for devices B and C. (a) Simulated partition coefficient $\mathcal{E}_{\rm mem}$ of devices~B~and~C. Data of device B (C) are red (blue). Data of each device are connected by lines to see the trend. Ideal calculated bandgaps are shown in grey. (b)-(f) Simulated displacement field for four kinds of modes. Color scheme represents the amplitude of displacement in a logarithmic scale. (b) An example of a non-membrane mode outside the bandgap. (c) An example of a MF mode inside the bandgap. (d) An example of a CF mode inside the bandgap. (e)-(f) Two examples of membrane modes inside/outside the bandgap.} \label{fig3}
\end{center}
\end{figure}

\begin{figure} \begin{center}
\includegraphics[width=0.5\textwidth]{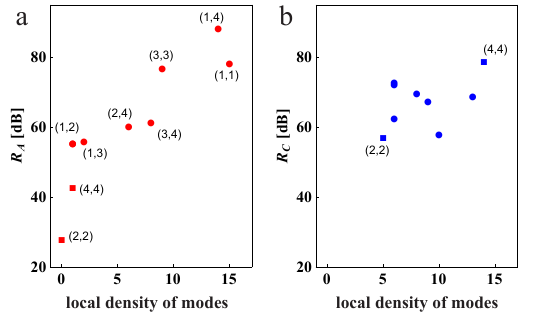}
\caption{The ratio of the actuated energy of the membrane modes provided by the piezoelectric actuator to that by the thermal fluctuating force as a function of a local density of modes. The density is determined from the data in Fig.~\ref{fig1}(c) by counting the number of observed modes in a 50~kHz range centered at each membrane mode. (a) Data for device A. The membrane mode indices are labeled. (b) Data for device C. The corresponding modes with lowest $R$ in (a) are shown in square.} \label{fig4}
\end{center}
\end{figure}

The observed eigenmodes include admixtures of modes created by the membrane, the MF, the PnC, and the CF. We use a FEM to simulate the whole device in order to visualize and characterize the expected frequency-dependent structure of all the modes. The boundary conditions for the simulation fix the corners of the back side of the chip. We find all the eigenmodes between 1~and~5~MHz.  To estimate the motion that will be observed on the \SiN~membrane [as measured in Fig.~\ref{fig2}(c) and (d)], for each mode we calculate a ``partition coefficient'' defined by the ratio of the energy stored in the membrane to the energy stored in the whole device
\begin{equation}
\mathcal{E}_{\rm mem} \equiv 
\frac{\int_{\rm mem} \varrho(\mathbf{x})|\mathbf{u}(\mathbf{x})|^2 d^3 x}{\int_{\rm whole} \varrho(\mathbf{x})|\mathbf{u(\mathbf{x})}|^2 d^3x},
\end{equation}
where $\mathbf{u}(\mathbf{x})$ is the simulated displacement field and $\varrho(\mathbf{x})$ is the mass density field.  

The partition coefficient $\mathcal{E}_{\rm mem}$ is plotted in Fig.~\ref{fig3} as a function of mode frequency using the parameters for devices B and C. The membrane modes are clearly identifiable as the $\mathcal{E}_{\rm mem}\simeq 0$~dB; these modes have the small effective mass associated with the \SiN~membrane.  A majority of the non-membrane modes of device C have an $\mathcal{E}_{\rm mem}$ between $-40$ to $-60$~dB; these modes have a much larger effective mass associated with the silicon substrate.   For device~B there are two ranges with reduced $\mathcal{E}_{\rm mem}$ that roughly overlap with the ideal calculated bandgaps.  The reductions are finite ($\mathcal{E}_{\rm mem}$ between $-70$ to $-130$~dB) and smoothly degraded because the simulation takes into account the finite number of unit cells.  There are also a finite number of non-membrane modes with $\mathcal{E}_{\rm mem} > -40$~dB.  Inside the device-B bandgap, these modes can be classified as defect modes with $\mathcal{E}_{\rm mem} < -30$~dB.   Outside of the device-B bandgap and in device C there is a larger number of modes ($\sim4\%$ of the modes) with $\mathcal{E}_{\rm mem} > -40$~dB.  These modes with the largest $\mathcal{E}_{\rm mem}$ tend to be clustered near the expected membrane mode frequencies.

In Fig.~\ref{fig3}(b)-(f) we also show the displacement profile of example modes on a logarithmic scale.  We see that the non-membrane modes inside the bandgaps are dominated by the MF or the CF, and the displacement field decays exponentially in the PnC [Fig.~\ref{fig3}(c),(d)]. On the contrary, the non-membrane modes outside the bandgaps have a uniformly distributed displacement field [Fig.~\ref{fig3}(b)]. We also find that for the membrane modes inside and outside the bandgaps [Fig.~\ref{fig3}(e) and (f)], the displacement fields in the PnC behave the same as the non-membrane modes inside and outside the bandgaps. In other words, inside the bandgap, the PnC acts as a passive mechanical filter that decouples the CF and the ``defect''; outside the bandgap, the PnC moves with all the other components together, i.e., they can be strongly coupled.

Lastly, we have studied the efficiency with which the piezoelectric actuator can drive membrane modes inside and outside the bandgap. The piezoelectric actuator does not directly drive the membrane; it drives the membrane through the chip frame, the PnC, and the membrane frame. In other words, the piezoelectric actuator actuates the membrane mode through the non-membrane modes, and hence we expect the driving efficiency to be low inside the observed bandgap.  We quantitatively analyze this effect by measuring the piezo actuated energy of the (1,1) through (4,4) membrane modes of devices A and C.  To obtain a calibrated measure of the relative actuated energy, we also measure for each mode the thermally actuated energy provided by the thermal fluctuating force, which is not shielded by the PnC.

In Fig.~\ref{fig4} we plot the ratio of the driven to thermal energies, $R$ (see Appendix) as a function of a measure that approximates the local mode density near each membrane mode. This measure provides an estimate of the expected driving efficiency, but not necessarily a rigorous correspondence, because the set of optically measured modes will not necessarily correspond to the set of modes that couples best to a particular membrane mode.   Nonetheless, we see a positive correlation between the driven motion and this mode density for device A.  We also observe a much larger dynamic range in $R$ for device A than for device C, which is as expected because the phononic crystal structure introduces a nonuniformity to the local mode structure.   A direct comparison between devices~A and C shows the smallest $R$ in device~A is 30~dB smaller than the smallest $R$ in device~C, indicating that in the bandgap membrane modes can be significantly isolated from the chip frame.

Delivering energy from the chip frame to the membrane is the reverse process of radiating energy from the membrane to the chip frame. Therefore, the well-isolated (small $R$) membrane modes are expected to have small radiation loss. However, the highest $Q$ of the membrane modes we observed in device A is about $10^6$, comparable with the highest $Q$ of the membrane modes in device C.  This is possibly because the membranes are still limited by the material loss from defects generated in this new fabrication process.  In the future, we will investigate realizing higher \SiN~$Qs$ in the bandgaps by measuring at cryogenic temperature and improving control of the fabrication.

We thank K. W. Lehnert, S. B. Papp, R. W. Andrews, and Y.-P.~Huang for valuable discussions, and J. A. Beall for helpful discussions on device fabrication. This work was supported by the DARPA QuASAR program, the ONR YIP, and the National Science Foundation under grant number 1125844. C.R. thanks the Clare Boothe Luce Foundation for support. P.-L.Y. thanks the Taiwan Ministry of Education for support.  This work is a contribution of the US Government; not subject to copyright.

\vspace{15pt}

\appendix*
\section{Appendix: Theory of the factor $R$}
To calibrate the actuated energy in membrane modes we compare to thermally driven energy via  the factor $R$. The actuated energy is proportional to the square of the vibration amplitude. Therefore we have:
\begin{align}
R=R(p,f,B_w)\propto\frac{|\mathcal{D}(p,f)/\eta|^2}{[\mathcal{S}_d(f)/\eta^2]B_w}=\frac{|\mathcal{D}(p,f)|^2}{\mathcal{S}_d(f)B_w},
\end{align}
where $\mathcal{D}(p,f)$ is the driven displacement amplitude measured with a network analyzer under external driving power~$p$, $\mathcal{S}_d(f)$ is the displacement spectral density measured with a spectrum analyzer without external driving power, $\eta$ is the overlap factor between the optical spot and the membrane mode shape, and $B_w$=2~Hz is the resolution bandwidth of the spectrum analyzer.

\end{document}